\documentclass[runningheads]{llncs}

\usepackage{amsmath}
\usepackage{algorithmic}
\usepackage{graphicx}
\usepackage{wasysym}
\usepackage{textcomp}
\usepackage{xcolor}
\usepackage{caption}
\usepackage{subcaption}
\usepackage{changepage}
\usepackage{url}
\usepackage{hyperref}
\usepackage{alltt}
\usepackage{balance}
\usepackage{booktabs}
\usepackage{soul}
\usepackage{float}
\usepackage{multirow}
\usepackage{wrapfig}
\usepackage{scalerel,amssymb}

\begin{document}

\title{Car drivers' privacy concerns and trust perceptions}

\author{Giampaolo Bella\inst{1}\and Pietro Biondi\inst{1}\and Giuseppe Tudisco\inst{1}}
\authorrunning{Bella et al.}
%
\institute{Dipartimento di Matematica e Informatica\\ Universit\`{a} degli Studi di Catania, Catania, Italy \\ \email{giamp@dmi.unict.it} \\ \email{pietro.biondi@phd.unict.it} \\ \email{giuseppe.tudisco@studium.unict.it}}

\maketitle

\begin{abstract}
Modern cars are evolving in many ways. Technologies such as infotainment systems and companion mobile applications collect a  variety of personal data from drivers to enhance the user experience. This paper investigates the extent to which car drivers understand the implications for their privacy, including that car manufacturers must treat that data in compliance with the relevant regulations. It does so by distilling out drivers' concerns on privacy and relating them to their perceptions of trust on car cyber-security. A questionnaire is designed for such purposes to collect answers from a set of 1101 participants, so that the results are statistically relevant. In short, privacy concerns are modest, perhaps because there still is insufficient general awareness on the personal data that are involved, both for in-vehicle treatment and for transmission over the Internet. Trust perceptions on cyber-security are modest too (lower than those on car safety), a surprising contradiction to our research hypothesis that privacy concerns and trust perceptions on car cyber-security are opponent. We interpret this as a clear demand for information and awareness-building campaigns for car drivers, as well as for technical cyber-security and privacy measures that are truly considerate of the human factor.
\end{abstract}

\section{Introduction}
The cars people are driving at present are complicated cyber-physical systems involving tight interaction between rapidly evolving car technologies and their human users, the drivers. 
To meet the needs and preferences of (at least) drivers, the infotainment system is more and more integrated with the setups for passengers' physical preferences, such as seating configuration, driving style and air conditioning, as well as for non-physical preferences, such as  music to play, preferred numbers to call and on-line payment details.

A plethora of data originates, whose processing enhances the driving experience and exceeds that towards increased support for autonomous driving, a goal of large interest at present. 
Modern cars also come with Internet connectivity ensuring, at least, that car software always gets  over-the-air updates from the manufacturer. Cars expose services remotely via dedicated apps that the driver installs on their smartphone to remotely operate functions such as electric doors, air conditioning, headlights, horn and even start/stop the engine. Therefore, car and driver's smartphone apps form a combined system that exposes innovative services, including locating the car remotely via GPS or even geo-fencing it, so that the app user would be notified if their car ever exceeds a predefined geographical area~\cite{connectedcarfeatures}.
Because cars are progressively resembling computers, offering services while treating personal data, they also attract various malicious aims.

The field of car cyber-security shows that software vulnerabilities can be exploited on a Jeep~\cite{remoteattackJeep}, on a General Motors~\cite{GM2015} as well as on a Tesla Model S~\cite{Tesla2016}. Such vulnerabilities may, in particular, impact data protection, and
the sequel of this manuscript will discuss the variety of personal data treated through cars, thus calling for compliance, at least in the EU, with the General Data Protection Regulation (GDPR)~\cite{gdpr}. Car cyber-security (``security'', in brief) is certain to be more modern than car safety, hence our overarching research goal is to understand whether the former is understood as well as the former is.
We formulate the hypothesis that privacy concerns decrease when trust perceptions on the underlying security and data protection measures are correspondingly high. For example, it means that if a driver feels that their personal data is protected, then that is because the driver trusts that the car is secure.
To assess such hypothesis, this paper does not take a common attack-then-fix approach but, rather, addresses the following research questions pivoted on drivers' perceptions. 
\begin{description}
\item[RQ1.] Are drivers adequately concerned about the privacy risks associated with how that their car and its manufacturer treat their personal data? 
\item[RQ2.] Do drivers adequately perceive the trustworthiness of their car, in terms of security especially?
\end{description}
We are aware that these research questions are not conclusive, and we have gathered data to specialise the answers by categories of drivers, e.g. by age or education. 
To the best of our knowledge, this is the first large-scale study targeting and relating privacy concerns and trust perception of car drivers. We took the approach of questionnaire development and survey execution through a crowdsourcing platform. Our goal was to get at least 1037 sets of responses in order for the findings to be statistically relevant, as explained below. We first piloted the questionnaire with 88 friends and colleagues with the aim of getting feedback but no significant tuning was required. After crowdsourcing, a total number of 1101 worldwide participants was reached.

We analysed the results obtained from the questionnaire through standard statistical analysis by Pearson's correlation coefficient, Spearman's rank correlation coefficient and Coefficient Phi. 
In a nutshell, most drivers believe that it is unnecessary for their car to collect their personal data because they find the collection unnecessary to the full functioning of modern cars; this indicates that privacy concerns are low, which in turn may be due to wrong preconceptions, given that cars do collect personal data.
Also, it appears that most drivers do not fully agree that their data is protected using appropriate security measures; this may be interpreted as a somewhat low trust on security. To our surprise, pairing these two abstracted findings clearly disproves our hypothesis.

Section~\ref{sec:related-work} comments on the related work,  Section~\ref{sec:method} outlines our research method, particularly the  questionnaire design, the crowdsourcing task and the statistical approach, Section~\ref{sec:results} discusses our results and Section~\ref{sec:conclusion} concludes.

\section{Related Work}\label{sec:related-work}
In 2014, Schoettle and Sivak~\cite{schoettle2014} surveyed public opinions in Australia, the United States and the United Kingdom regarding connected vehicles. 
Their research noted that people (drivers as well as non-drivers) expressed a high level of concern about the safety of connected cars, which does not seem surprising on the basis of the novelty of the concept at the time.
However, participants demonstrated an overall positive attitude towards connected car technology, with particular interest in device integration and in-vehicle Internet connectivity.
In 2016, Derikx et al.~\cite{derikx2016} investigated whether drivers' privacy concerns can be compensated by offering monetary benefits.
They analysed the case of usage-based auto insurance services where the rate is tailored to driving behaviour and measured mileage and found out that drivers were willing to give up their privacy when offered a small financial compensation.
Therefore, what appears to be missing is a study on drivers' understanding on the amount and type of personal data that modern cars process, which is the core of this paper. 

There are relevant publications on drivers' trust on car safety but are limited to self-driving cars. Notably, Du et al.~\cite{du2019} conducted an experiment to better understand whether explaining the actions of automated vehicles promote general acceptance by the drivers. They found out that the specific point in time when explanations were given was crucial for their effectiveness --- explanations provided before the vehicle started were associated with higher trust by the subjects.
Similar results were obtained by Petersen et al.~\cite{petersen2019} in another study in 2019. They manipulated drivers' situational awareness by providing them with different types and details of information. Their analysis showed that situational awareness influenced the level of trust in automated driving systems, allowing drivers to immerse themselves in non-driving activities. Clearly, the more people are aware of something, the more trust they manage to place in it.

It is clear that modern cars technologies are not limited to self-driving features. Modern cars include innumerable digital components, often integrated in the infotainment system, which interact with drivers and collect their data. It follows that modern cars process personal data to some extent, as detailed in the next Section, hence car manufacturers must meet specific sets of requirements to comply with the relevant regulations. Therefore, it becomes important to assess drivers' concerns on their privacy through their use of a car and drivers' trust on the security (also in relation to their trust on the safety) of the car.

\section{Research method}\label{sec:method}
We took the approach of questionnaire development and survey execution to assess car drivers' privacy concerns and trust perceptions.  Specifically, we built questionnaire with 10 questions, administered it through a crowdsourcing platform and carried out a statistical analysis of the answers. Opinions were measured using a standard 7-point Likert scale. With a very low margin of error, of just 4\%, and a very high confidence level, of 99\%, the necessary sample size to represent the worldwide population is 1037. Our total respondents were 1101, including piloting over 88, so our findings are statistically relevant of the entire world --- a limitation is that, while Prolific ensures that respondents are somehow geographically dispersed, it cannot guarantee that they are truly randomly sampled from the entire world.

\section{Results}\label{sec:results}
The answers are catalogued and statistically studied by analysing indexes of central tendency and correlation coefficients. The indexes of central tendency (mean and median) synthesise with a single numerical value the values assumed by the data. The mean value is coupled with the standard deviation in order to measure the amount of variation of the values. There is no room to present the demographic and its correlations with other answers here; we recall that driving at least 3 hours a week was a prerequisite to enter the study, along with being over 18. 

To simplify the analysis of the answers to the core questions, we follow the standard practice of grouping the 7 levels of agreement into three categories. Specifically, if the participants reply with ``Strongly agree'', ``Agree'' or ``Somewhat agree'', then we consider their value as ``Agreeing''; if instead they select ``Neither agree nor disagree'', then we consider them in the category ``Undecided''; and finally, if the participants select ``Somewhat disagree'', ``Disagree'' or ``Strongly disagree'', then we consider those answers as ``Disagreeing''.

\subsubsection{Knowledge on modern cars}
Question Q1 evaluates the driver's knowledge on modern cars. Considering the values of the mean and the median, shown in Table~\ref{tab:q1}, it can be stated that the interviewed sample considers itself knowledgeable about modern cars. The data show that 55\% of participants are quite confident about their knowledge, while a minority of the participants (about 29\%) think they are not. Finally, 16\% of participants think they have average knowledge about modern cars. Thus, considering the answers of the preliminary question, there does not seem to be a substantial difference between those who drive a few hours a week and those who drive more with regard to the level of knowledge they claim to have on modern cars.

Then, question Q2 asks respondents whether or not they agree that modern cars are similar to modern computers. Also this question receives a high rate of agreement. We note that 72\% of participants agree that a modern car is similar to a modern computer. Furthermore, it turns out that 14\% of them are undecided while 14\% of them disagree with the statement. The mean and the median are shown in Table~\ref{tab:q2}.

\begin{table}[ht]
\centering
\caption{Q1, Q2 answers and their statistics}
\label{tab:q1}
	\begin{tabular}{lc}
		\toprule
		\textbf{Knowledge level}            & \textbf{[\%]} \\
		\midrule
		Knowledgeable about modern cars     & 55            \\
		Average knowledge                   & 16            \\
		Not knowledgeable about modern cars & 29            \\
		\midrule
		\midrule
		Mean                                & 4.37          \\
		Median                              & 5             \\
		Standard Deviation                  & 1.55          \\
		\bottomrule
	\end{tabular}
\quad
\label{tab:q2}
	\begin{tabular}{lc}
		\toprule
		\textbf{Agreement level} & \textbf{[\%]} \\
		\midrule
		Agreeing                 & 72            \\
		Disagreeing              & 14            \\
		Undecided                & 14            \\
		\midrule
		\midrule
		Mean                     & 5             \\
		Median                   & 5             \\
		Standard Deviation       & 1.35          \\
		\bottomrule
	\end{tabular}
\end{table}

\subsubsection{Concerns on data privacy}
The first of these questions (Q3) asks participants to select all the categories of data they think a car collects. It must be remarked that this answer allows for multiple choice, so a respondents can choose from multiple categories of data. Table~\ref{tab:q3} shows the answers selected by the respondents. The predominant categories according to the interviewed sample are: ``personal data about the driver'' (selected by 56\% of the sample); ``public data about the driver'' (selected by 54\% of the sample); ``public data not about the driver'' (selected by 47\% of the sample). 
A few participants think that their vehicle collects more sensitive data belonging to the special categories of personal (13\%) and financial data (11\%). Finally, we note that just 8\% of the participants think that modern cars do not collect any data at all. 

Overall, these findings confirm a modest level of awareness in terms of what data a car collects. In particular, while it is positive that the majority (56\%) understands that driver's personal data are involved, it is concerning that a similar subset (54\%) deem such data about the driver to have been made public. It would be surprising if any car manufacturer's privacy policy stated that the driver's collected data would be made public (and such policies are well worthy of a dedicated comparative study). This potential confusion calls for awareness campaigns, more readability of official documents and innovative technologies to ensure policies are understood. By contrast, a positive sign that a small kernel of participants is highly informed is the appreciable understanding that special categories of personal data (13\%) or financial data (11\%) may be gathered. 

\begin{table}[ht]
	\centering
	\caption{Q3 answers}\label{tab:q3}
	\scalebox{0.95}{
		\begin{tabular}{lc}
			\toprule
			\textbf{Collected data}                              & \textbf{[\%]} \\
			\midrule
			Personal data about the driver                       & 56            \\
			Public data about the driver                         & 54            \\
			Public data not about the driver                     & 47            \\
			Special categories of personal data about the driver & 13            \\
			Financial data about the driver                      & 11            \\
			No data at all                                       & 8             \\
			\bottomrule
	\end{tabular}}
\end{table}

Question Q4 asks participants whether they think it is necessary to collect personal data to achieve full vehicle functionality. The indexes and a summary of the answers are shown in Table~\ref{tab:q4}. It shows that 27\% of the participants agree with the statement above, moreover, 19\% of them are undecided and 54\% of them disagree with the statement. Thus, we could argue that the participants disagree with the statement proposed in the question. 

This finding can be interpreted in various ways. On one hand, it denounces a false preconception because that the customised, driver-tailored experience that is getting more and more common at present is certain to stand on a trail of data collected about the driver's.
It clearly also signifies that drivers are neither adequately informed on what data is being collected and for what purposes, contradicting art. 5 of GDPR, nor have they been able to grant an informed consent, contradicting art. 7 of GDPR.

\begin{table}[ht]
\centering
\caption{Q4, Q5 answers and their statistics}
\label{tab:q4}
	\begin{tabular}{lc}
		\toprule
		\textbf{Agreement level} & \textbf{[\%]} \\
		\midrule
		Agreeing                 & 27            \\
		Disagreeing              & 54            \\
		Undecided                & 19            \\
		\midrule
		\midrule
		Mean                     & 3.35          \\
		Median                   & 3             \\
		Standard Deviation       & 1.58          \\
		\bottomrule
	\end{tabular}
\quad
\label{tab:q5}
\begin{tabular}{lc}
		\toprule
		\textbf{Agreement level} & \textbf{[\%]} \\
		\midrule
		Agreeing                 & 21            \\
		Disagreeing              & 65            \\
		Undecided                & 14            \\
		\midrule
		\midrule
		Mean                     & 2.97          \\
		Median                   & 3             \\
		Standard Deviation       & 1.67          \\
		\bottomrule
	\end{tabular}
\end{table}

Moving on to the answers of question Q5, it can be noticed that just 21\% of the sample agrees to the transmission of data over the Internet, only 14\% of participants are undecided moreover 65\% of them disagree with the statement. This means that the sample is not very convinced to send personal data over the Internet. Table~\ref{tab:q5} shows agreement levels and indexes of Q5's answers. 

This may again be interpreted as a wrong preconception because it is clear that remote services, including eCall, location-tailored weather forecasts, music streaming and many others, must generate Internet traffic.

\subsubsection{Perceptions of trust on safety}
Question Q6 asks whether participants agree that a modern vehicle safeguards the life of its driver. The agreement levels and the indexes of central tendency are shown in Table~\ref{tab:q6}. It turns out that 77\% of participants agree with the statement above, 
then just 8\% disagree with the statement, and 15\% of them are undecided.

Question Q7 asks participants whether a modern car protects its driver's personal data better than its driver's life. It appears that a part of the sample is undecided with this statement (26\%), just 18\% of participants agree with the statement moreover 56\% of them disagree. Table~\ref{tab:q7} shows also that the indexes of central tendency are not as high when compared to the previous question.

There is considerable uncertainty in front of this question, if not for the majority's expression of disagreement (56\%). It signifies that trust on security still has a great lot to grow in comparison to trust on safety, perhaps due to the much longer establishment of the latter. It is well known that trust may take a long time to root, and car security is certain to be a somewhat recent problem.

\begin{table}[ht]
\centering
\caption{Q6, Q7 answers and their statistics}
\label{tab:q6}
	\begin{tabular}{lc}
		\toprule
		\textbf{Agreement level} & \textbf{[\%]} \\
		\midrule
		Agreeing                 & 77            \\
		Disagreeing              & 8             \\
		Undecided                & 15            \\
		\midrule
		\midrule
		Mean                     & 5.26          \\
		Median                   & 5             \\
		Standard Deviation       & 1.20          \\
		\bottomrule
	\end{tabular}
\quad
\label{tab:q7}
	\begin{tabular}{lc}
		\toprule
		\textbf{Agreement level} & \textbf{[\%]} \\
		\midrule
		Agreeing                 & 18            \\
		Disagreeing              & 56            \\
		Undecided                & 26            \\
		\midrule
		\midrule
		Mean                     & 3.26          \\
		Median                   & 4             \\
		Standard Deviation       & 1.46          \\
		\bottomrule
	\end{tabular}
\end{table}

\subsubsection{Perceptions of trust on security}
Question Q8 asks whether the data collected from the vehicle is legitimately processed according to the relevant regulations. Table~\ref{tab:q8} shows that 44\% of the participants agree with this statement moreover 25\% disagree and the rest of them (31\%) are undecided.

Trust one the legitimacy of the data processing is not higher than 44\%. This indicates, once more, that car drivers need to be better informed, first of all. Conversely, this means that the majority, 56\%, are not sure about the legitimacy of the processing of their personal data. Being informed correctly is essential for raising awareness, which in turn is essential for trust building.

Question Q9 asks if participants believe that the personal data collected is systematically analysed and evaluated using automated processes (including profiling). From Table~\ref{tab:q9}, around 42\% of participants agree with this statement, moreover 32\% of them disagree and 26\% are undecided with the statement. 

This question is designed to be self-contained and understandable by everyone.
A notable 42\% show concern that profiling takes place, which may be taken to signify a correspondingly low trust on the security of the treatment. There is no official public information on whether car manufacturers really carry out profiling but, if this were the case, then a Data Protection Impact Assessment, pursuant art. 35 of GDPR, would have been necessary.

The last question (Q10) asks whether the participants feel that the data transmitted over the Internet are protected by adequate technologies. Table~\ref{tab:q10} confirms the representation that agrees with the question (46\%) to be considerable.
The fact that those who agree do not exceed the majority of the sample clearly indicate, also in this case, room for improving drivers' trust on security.

\begin{table}[ht]
\centering
\caption{Q8, Q9, Q10 answers and their statistics}
\label{tab:q8}
	\begin{tabular}{lc}
		\toprule
		\textbf{Agreement level} & \textbf{[\%]} \\
		\midrule
		Agreeing                 & 44            \\
		Disagreeing              & 25            \\
		Undecided                & 31            \\
		\midrule
		\midrule
		Mean                     & 4.28          \\
		Median                   & 4             \\
		Standard Deviation       & 1.31          \\
		\bottomrule
	\end{tabular}
\quad
\label{tab:q9}
	\begin{tabular}{lc}
		\toprule
		\textbf{Agreement level} & \textbf{[\%]} \\
		\midrule
		Agreeing                 & 42            \\
		Disagreeing              & 32            \\
		Undecided                & 26            \\
		\midrule
		\midrule
		Mean                     & 4.07          \\
		Median                   & 4             \\
		Standard Deviation       & 1.43          \\
		\bottomrule
	\end{tabular}
\quad
\label{tab:q10}
	\begin{tabular}{lc}
		\toprule
		\textbf{Agreement level} & \textbf{[\%]} \\
		\midrule
		Agreeing                 & 46            \\
		Disagreeing              & 32            \\
		Undecided                & 22            \\
		\midrule
		\midrule
		Mean                     & 4.19          \\
		Median                   & 4             \\
		Standard Deviation       & 1.49          \\
		\bottomrule
	\end{tabular}
\end{table}

\subsection{Correlations}
There are statistically significant correlations that arise by analysing the relevant coefficients over the data obtained from the sample. In brief,
Pearson's linear correlation coefficient, denoted by the letter $r$, allows us to evaluate a possible linearity relationship between two sets of data.
Spearman's rank correlation coefficient, denoted by the Greek letter \( \rho \),  measures the correlation between two numerical variables; these must be sortable because Spearman's correlation coefficient is defined as Pearson's correlation coefficient applied to the ranks.
The Phi coefficient (or mean square contingency coefficient), denoted with the Greek letter \( \phi \), is a measure of association for two binary variables and is calculated from the frequency distributions of the pairs.
Correlation coefficient values are accompanied by a significance level (the \textit{p-value}) to establish the reliability of the calculated value. The p-value is a number between 0 and 1 representing the probability that the result would have been obtained if the data had not been correlated. If the p-value is less than 0.01 then the relationship found is statistically significant.

The analysis looks at the core questions, those from Q1 to Q10, to focus on general correlations between knowledge on the subject matter, privacy concerns, trust perceptions of safety and on security. 

\subsubsection{Core question analysis}
We noted a significant correlation between question Q1 and question Q2 (\(\rho = 0.48\), \(p < 0.001\)). Therefore, it seems that participants who are knowledgeable about modern cars also think that modern cars are similar to modern computers, reinforcing the conclusion. Moreover, thanks to the correlation between question Q1 and question Q4 (\(\rho = 0.35\), \(p < 0.001\)) we can state that those who consider themselves informed about modern cars also believe that the data collected by the car is necessary for the full functioning of the car. This aligned with our own, specialist view. There is also a significant correlation between question Q1 and question Q6 (\(\rho = 0.40\), \(p < 0.01\)), that is, those who are knowledgeable about modern cars think that a modern car safeguards its driver's life. Somewhat surprisingly, it appears that Q1 does not significantly correlate with later questions on trust on car security, signifying that trust on security must grow even for those who are knowledgeable about the field.

We calculated the Phi coefficients between the answers of question Q3 to determine if there are any associations, i.e. whether there are pairs of categories of personal data that appear together in the answers. The coefficient values are shown in Table~\ref{tab:corQ3}, and it becomes apparent that there are only two values that may establish a possible association. The Phi Coefficient obtained between the couple ``Special categories of personal data'' and ``Financial data about the driver'' is 0.3255, which means that those who think that financial data are collected by modern cars also think that special categories of personal data are collected as well. This exhibits a correct preconception because financial data are routinely grouped with special categories of data. Also, given the 0.3363 value, we notice that drivers who think ``Special categories of personal data'' are collected from the car, also think that ``Personal data about the driver'' are collected, emphasising a correct understanding that personal data also include the special categories (of personal data).

\begin{table}[ht]
	\centering
	\caption{Phi Coefficients of question 3}\label{tab:corQ3}
	\scalebox{0.84}{
		\begin{tabular}{c|cccccc}
			\( \phi \)       & No Data & Fin             & Spec            & Pub     & Pub\(_{driver}\) & Pers \\
			\hline
			No Data          & 1       &                 &                 &         &                  &      \\
			Fin              & -0.0920 & 1               &                 &         &                  &      \\
			Spec             & -0.0999 & \textbf{0.3255} & 1               &         &                  &      \\
			Pub              & -0.2371 & 0.0004          & -0.0624         & 1       &                  &      \\
			Pub\(_{driver}\) & -0.2973 & 0.1468          & 0.0759          & 0.0255  & 1                &      \\
			Pers             & -0.3136 & 0.2332          & \textbf{0.3363} & -0.2099 & 0.1743           & 1    \\
	\end{tabular}}
\end{table}

Moving on to question Q4, there is a strong statistically significant correlation between question Q4 and question Q5: both the Pearson coefficient and the Spearman coefficient have very high values (\(r = 0.55\), \(\rho = 0.68\)) both with a reliability value \(p < 0.001\). In consequence, we can affirm that those who think that it is necessary to collect personal data for the full functioning of their vehicle also think that this data should be transmitted over the Internet. 
There is also another statistically significant correlation between question Q4 and question Q8 (\(r = 0.39\), \(\rho = 0.50\), \(p < 0.01\) for both), showing that those who agree to the collection of personal data also think that the data are processed legitimately in a manner consistent with the relevant regulations. Both of these can be taken as indications that those with modest privacy concerns show some trust on security, but we are mindful of the generally low agreement with Q4 and Q5 and only fair agreement with Q8 noted above.

Question Q5 correlates only moderately with question Q8 (\(\rho = 0.48\), \(p < 0.01\)) but more significantly with question Q10 (\(r = 0.36\), \(\rho = 0.52\), \(p < 0.01\)). It follows that those who think that data should be transmitted over the Internet, also think that this data will be adequately protected during transmission. This shows that trust on security is broad if present.

Spearman's correlation coefficient detects a moderately significant correlation between question Q6 and question Q8 (\(\rho = 0.45\), \(p < 0.01\)), so that it seems that those who think that a modern car safeguards its driver's life also think that the personal data collected are processed legitimately according to the relevant regulations in force. This seems a positive outcome in terms of a spreading of trust on safety over trust on security. It is unfortunate that this correlation is not very strong, and we deem it highly desirable to develop socio-technical security and privacy measures to reinforce it in the future.

There is a statistically significant correlation between question Q7 and question Q10 (\(r = 0.38\), \(\rho = 0.52\), \(p < 0.01\)).
In fact, those who think that a modern car protects its driver's personal data better than it safeguards its driver's life also think that the personal data are protected by adequate technology when the vehicle transmits it over the Internet. The Spearman's correlation coefficient also shows a significant correlation between question Q7 and question Q8 (\(\rho = 0.47\), \(p < 0.01\)), that is, those who think that a modern car protects its driver's data better than it safeguards its driver's life also think that the personal data collected are processed legitimately according to the relevant regulations. These findings confirm that trust on security is somehow ``logical'' in the sense that it covers all relevant elements. 

There is also a significant correlation between question Q8 and question Q9 (\(\rho = 0.41\), \(p < 0.01\)), it appears that those who think that modern cars carry out a systematic and extensive evaluation of personal data also think that their data are processed in a legitimate way according to relevant regulations. This correlation suggests that drivers who consent to the evaluation of their personal data even consent to profiling --- perhaps too lightheartedly, raising concern that the potentially negative consequences of profiling may not be fully understood at present. It may be inferred that drivers are not fully aware that it would be their right to object to profiling, as prescribed by art. 22 of GDPR.

We also noted a moderate correlation between question Q9 and question Q4 (\(\rho = 0.45\), \(p < 0.01\)). Those who think that in order to use the full functionality of the car it is necessary to provide personal data also think that this data is analysed and studied according to automatic processes to evaluate personal aspects of drivers. This reconfirms that profiling is somewhat ill-understood. There is also a statistically significant correlation between question Q9 and question Q5 (\(\rho = 0.46\), \(p < 0.01\)) indicating that those who think that their data are analysed by automatic evaluation processes also think that they are transmitted over the Internet. This outcome correctly indicates that potential profiling does not take place aboard the car.

There is a statistically significant correlation between question Q10 and question Q4 (\(r = 0.37\), \(\rho = 0.49\), \(p < 0.01\)), that is, those who think that the personal data collected by the vehicle is necessary for the full functioning of the car also think that their data is adequately protected when transmitted over the Internet. Once more, modest privacy concerns lead to some trust on security. Finally, there is a statistically significant correlation between question Q10 and question Q8 (\(r = 0.51\), \(\rho = 0.64\), \(p < 0.01\)), so we can argue that those who think that their personal data is processed lawfully also think that the data are adequately protected over the Internet. Here is yet another confirmation that trust on security, if at all present, covers all relevant aspects.

\section{Conclusions}\label{sec:conclusion}
Our study was designed with care to carve out drivers' privacy concerns and trust perceptions with the ultimate aim of assessing our research hypothesis that low privacy concerns imply high trust perceptions. Crowdsourcing was leveraged to collect a representative sample of participants. Answers were then analysed in isolation as well as statistically correlated, producing very many insights.
There would be little use in developing amazing technical security and privacy measures for preserving drivers' privacy and the security of their cars in case drivers
are not adequately concerned about
the privacy issues bound to their driving and yet do not trust the security of their cars at an appropriate level. That case is confirmed by the results of our study, thus contradicting our research hypothesis. We consider this outcome worthy of further attention.

Precisely, we believe that the privacy concerns that arose are insufficient in the present technological setting. We would have found it more positive if drivers exhibited higher awareness on the personal data involved through their driving, on how treating such data is fundamental for delivering driver-tailored services, and on the fact that such service quality often demands data transmission over the Internet. Unfortunately, the opposite scenario holds. 
A somewhat logical explanation of low privacy concerns could be a high trust on security, but we were surprised once more that also trust on security was somewhat low. Therefore, the only way to read the general outcome is that privacy is generally ill-understood by drivers, hence we learn that more information must be delivered to them in order to raise awareness and then form correct privacy concerns and correspondingly adequate trust perceptions. 
We strongly argue that this must be the ultimate effect for the development of more and more advanced technical security and privacy measures.

The correlations among answers could be seen as somewhat logical.
For example, knowledge on the field correlates with adequate privacy concerns and well-related trust perceptions. It is noteworthy that the potentially negative implications of profiling on the freedoms of natural persons are far from being well received at the moment. 
Trust on security is much less represented than trust on safety, arguably because the former derives from a less rooted perception in our society due to the relatively young age of the technologies that should support it. 
Moreover, trust on cyber-security is normally broad, that is, if it is present to some extent, it then covers all relevant aspects. 
We ultimately maintain that also correlations justify a need for more awareness and trust building campaigns.

The value of our results is multifaceted. They can be read in support of the ISO/SAE DIS 21434 standard, which is yet to be finalised. They also offer a solid baseline to conduct a cyber-security and privacy risk assessment on cars following standard methodologies such as ISO/IEC 27005.
Future work includes tailoring the effort presented in this paper to specific car brands in support of a contrastive analysis among brands. It is clear that the user-level studies in the automotive field that this paper incepted have great potential for growth.

\paragraph{Acknowledgments}
This research was funded by COSCA (COnceptualising Secure Cars)~\cite{COSCA}, a project supported by the European Union's Horizon 2020 research and innovation programme under the NGI TRUST grant agreement no 825618.

\bibliographystyle{splncs04}
\bibliography{references}

\end{document}